\documentclass{IAU-TrA}
\usepackage{graphicx}

\title[VARIABLE STARS]     
{}

\author[DIVISION~V / COMMISSION 27]   
{}

\pubyear{2012}
\volume{Volume XXVIIIA}
\pagerange{001--008}
\jname{Transactions IAU, Volume XXVIIIA}
\editors{Ian Corbett, ed.}
\begin{document}

\maketitle

{\bf

\large
\begin{tabbing}
\hspace*{65mm}       \=                                              \kill
COMMISSION~27         \> VARIABLE STARS                            \\
                     \> {\small\it \a'{E}TOILES VARIABLES}         \\
\end{tabbing}

\normalsize

\begin{tabbing}
\hspace*{65mm}       \=                                              \kill
PRESIDENT            \> Gerald Handler                               \\
VICE-PRESIDENT       \> Karen R. Pollard                             \\
PAST PRESIDENT       \> Steven Kawaler                         \\
ORGANIZING COMMITTEE \> Margarida S. Cunha, Katalin Olah,\\
                     \> Katrien Kolenberg, C. Simon Jeffery,\\
                     \> M\a'{a}rcio Catelan, Laurent Eyer, \\
                     \> Timothy R. Bedding, S. O. Kepler,         \\
                     \> David Mkrtichian            \\   
\end{tabbing}

\bigskip

\noindent
TRIENNIAL REPORT 2009-2012
}

\firstsection 

\section{Introduction}

As research on variable stars continues at an ever growing pace, this 
report can only give a selection of research highlights from the past 
three years, with a rigorously abbreviated bibliography. The past 
triennium has been dominated by results of the CoRoT (Astronomy \& 
Astrophysics 2009) and {\em Kepler} (Gilliland et al.~2010) space missions,
stemming from their unprecedented photometric accuracies and large time bases.

Numerous conferences related to the field have taken place; the 
following list certainly is incomplete. Two of the biennial 
meetings on stellar pulsation were held in Santa Fe, USA (2009) and 
Granada, Spain (2011). Kepler Asteroseismic Science Consortium Workshops 
took place in Aarhus, Denmark (2010) and Boulder, USA (2011); CoRoT 
symposia in Paris, France (2009) and Marseille, France (2011). The 
Kukarkin Centenary Conference was held in Zvenigorod, Russia (2009) and 
IAU Symposium 272 on active OB stars in Paris (France, 2010). IAU 
Symposium 273 dealed with the Physics of Sun and Star Spots (Ventura, 
USA, 2010), IAU Symposium 286 with Comparative Magnetic Minima 
(Mendoza, Argentina, 2011). IAU Symposium 285 (New Horizons in Time 
Domain Astronomy, Oxford, UK, 2011) had substantial interest for
C27. The American Association of Variable Star Observers held its 
$100^{\rm th}$ annual meeting 2011 in Boston, USA.

\section{Science highlights}

\subsection{Stellar activity}

A comprehensive review of the origin and properties of starspots was
published by Strassmeier (\cite{strassmeier}), with special emphasis
on the the possibility of detecting exoplanets around spotted host
stars.  The existence of spots for discovering planets and in studying
possible star-planet interactions has recently been recognised.

Several papers have been devoted to the active star CoRoT-7 and its
planet(s). The review by Pont et al.\ (2011) highlights the
importance of a realistic treatment of both the activity and the
uncertainties which are applicable to most small-planet candidates
observed by CoRoT and {\em Kepler}. The combined presence of activity
and additional errors precludes a meaningful search for additional
low-mass companions.

Deming et al.~(\cite{deming}) developed a method to correct planetary
radii for the presence of both crossed and uncrossed star spots. The
exo-Neptune HAT-P-11b transits nearly perpendicular to the stellar
equator, and the authors related the dominant phases of star-spot
crossings to active latitudes on the star. Precise transit
measurements over long durations may allow one to construct a butterfly
diagram to probe the cyclic evolution of magnetic activity on the
active K-dwarf planet-host star.

Osten et al.~(\cite{osten}) observed a large stellar flare and 
fluorescence from the dMe star EV Lac. The size of the flare, in terms 
of its peak X-ray luminosity, exceeded the non-flaring stellar bolometric 
luminosity, providing important constraints on the time-scales for 
energy storage and release in a stellar context.

Koll\'ath \& Ol\'ah (\cite{kollath}) and Ol\'ah et al.\ (\cite{olah}) 
demonstrated that time-frequency distributions provide useful tools for 
analysing the observations of active stars whose magnetic activity 
varies with time. Their technique applied to sunspot data revealed a 
complicated, multi-scale evolution in solar activity. Time 
variations in the cycles of 20 active stars based on decade-long 
photometric or spectroscopic observations show that stellar activity 
cycles are generally multiple and variable (see also Sect.\ 2.2).

Korhonen et al.~(\cite{korhonen}) presented simultaneous low-resolution
longitudinal magnetic field measurements and high-resolution
spectroscopic observations of the cool single giant FK~Com.  The
maxima and minima in the mean longitudinal magnetic field are both
detected close to the phases where cool spots appear on the stellar
surface.

\subsection{Solar-like oscillations}

Oscillations in the Sun are excited stochastically by convection, as also
they are in other stars with convective envelopes.  CoRoT has contributed 
substantially to the list of main-sequence and subgiant stars with 
solar-like oscillations (e.g.~Mathur et al.\ 2010 and references 
therein). A problem particularly present in F-stars (e.g.~Benomar et 
al.\ 2009) was illuminated by CoRoT data: the short-mode lifetimes cause 
blending of some oscillation modes, thereby hampering mode identification.

A major achievement by CoRoT came from observations of hundreds of G- and
K-type red giants showing clear oscillation spectra that are
remarkably solar-like, with both radial and non-radial modes
(e.g.~De~Ridder et al.\ 2009, Mosser et al.\ 2011). CoRoT continues to
produce excellent results on red giants, main-sequence and subgiant
stars.

The {\em Kepler} mission carried out a survey targeting more than 2000
main-sequence and subgiant stars for one month each. Solar-like
oscillations were detected, and clear measurements of the large
frequency separation for about 500 stars were made (Chaplin et
al.~2011) -- an increase by a factor of $\sim$20 over previous
results.  Some of these stars have been studied individually. The first
results on two main-sequence stars and one subgiant provided evidence
for mixed-mode oscillations in the latter (Chaplin et al.\ 2010).

{\em Kepler} detected solar-like oscillations in thousands
of red giants (e.g.\ Hekker et al.\ 2011 and references therein),
including some in open clusters (e.g.\ Basu et al.\ 2011), enabling
asteroseismic studies that would not have been thought possible
only a few years ago. For example, the gravity-mode period spacings in
red giants provide a means to disentangle the evolutionary phases of
hydrogen- and helium-burning in red-giant stars (Bedding et
al.\ 2011), an otherwise difficult task given the similarities in the
mass, luminosity and radius of these two groups.  Miglio et al.\ (2010)
described the first detection of the seismic signature of the helium
second-ionization region in red-giant stars, opening up the interesting
possibility of determining seismically the helium content of their
envelopes.

Garc{\'{\i}}a et al.~(2010) presented the first strong evidence for
cyclic frequency variations associated with the presence of a stellar
magnetic-activity cycle in a star other than the Sun.  Seismic
signatures of stellar activity cycles, in combination with additional
information such as differential rotation, extent of convective
envelope, etc., have the potential to increase substantially our
understanding of mechanisms for magnetic-field generation and
evolution.  In the solar case, Fletcher et al.~(2010) hypothesized a
second dynamo based on quasi-biennial solar oscillation frequency
variations.

Solar-like and heat-engine oscillations are not mutually
exclusive. Both may in fact operate in a star provided that the
surface convection layer is thin (Samadi et al.\ 2002).  There is
evidence from {\em Kepler} data for solar-like oscillations in at
least one $\delta$~Scuti star (Antoci et al.\ 2011). There has even
been a suggestion that sub-surface convection in B-type stars could
excite solar-like oscillations (Cantiello et al.\ 2009, Belkacem et
al.\ 2010). Meanwhile, Degroote et al.~(2010) suggested that
stochastically-excited oscillations are revealed in CoRoT photometry
of the O-type star HD~46149.

\subsection{Classical and heat-driven main sequence pulsators}

The enigmatic Blazhko effect (amplitude/phase modulation) turns out to 
be very common in RR Lyrae stars, as ground-based (Jurcsik et al.~2009), 
CoRoT (Szab\'o et al.~2010a), and {\em Kepler} results confirm 
(e.g.~Kolenberg et al.~2010). The pulsations in at least some RR 
Lyrae stars are remarkably stable (Nemec et al.~2011).

Space photometry data of RR Lyrae stars reveal previously unseen
features in Blazhko stars, such as period doubling (Kolenberg et
al.~2010) and additional modulations, which are not fully understood
yet (Benk\H{o} et al.~2010, Guggenberger et al.~2011).  This is also
the case for RR Lyr itself, successfully observed by {\em Kepler}
(Kolenberg et al.\ 2011). Period doubling has been traced to a 9:2
resonance between the fundamental mode and the ninth radial overtone
(Szab\'o et al.~2010b), confirmed by models of Buchler \& Koll\'ath
(2011). Their results may even explain the irregular amplitude modulation
which recent observations reveal.  At last we may come closer to an
explanation for the Blazhko effect.

RR Lyrae stars continue to fulfill their role as indicators, not only
of distance but also of tracers of galaxy formation histories. They
are increasingly being used by large-scale surveys such as SDSS
(e.g.~Sesar et al.~2010).  In particular, studies of the so-called
Oosterhoff dichotomy, which until recently were confined to the Milky
Way and its nearest neighbours (e.g.~Catelan 2009), can now probe
greater distances (e.g., Fiorentino et al.~2010), and even include
globular clusters as far away as M31 (Clementini et al.~2009).

Engle et al.\ (2009) reported X-ray emission of three bright Cepheids 
observed with XMM-Newton and Chandra. Despite differences in spectral 
type and pulsation properties, the Cepheids have similar X-ray 
luminosities and soft-energy distributions. Such high energy could arise 
from warm winds, shocks or pulsationally-induced magnetic activity.

Herschel images of Mira (Mayer et al.\ 2011) reveal broken arcs and
faint filaments in the ejected material of the primary star.  Mira's
IR environment appears to be shaped by the complex interaction of its
wind with its companion, the bipolar jet and the ISM.  High-angular
resolution Chandra imaging by Karovska et al.\ (2011) indicated
focused-wind mass accretion, a ``bridge" between Mira A and Mira B,
indicating gravitational focusing of the Mira A wind whereby
components exchange matter directly as well as by wind accretion.
That greatly helps explain accretion processes in symbiotic systems
and other detached and semi-detached interacting systems.

The ``Cepheid Mass Problem" (the mismatch between masses computed from
evolutionary tracks and hydrodynamic pulsation calculations for
classical Cepheids) is a fundamental test of stellar-evolution
models. The problem may be related to convective overshoot and
possible mass loss (e.g.~Neilson et al.~2011).  Marengo et al.~(2010)
discovered an infrared nebula and a bow shock around $\delta$ Cephei
and its hot companion, supporting the hypothesis that $\delta$ Cephei
may be currently losing mass.

The {\em Kepler} characterization of the variability in A- and F-type stars 
(Grigahc\`ene et al.~2010, Uytterhoeven et al.~2011) revealed a large 
number of hybrid $\delta$ Sct/$\gamma$ Dor pulsators, thereby opening up an 
exciting new channel for asteroseismic studies.

Intriguingly, Kurtz et al.~(2011) presented the first example of a
star that oscillates around multiple pulsation axes.  Evidence for the
presence of torsional modes was also given (also a first). Sousa \&
Cunha~(2011) gave the first theoretically-based explanation for the
diversity found observationally in the atmospheric behaviour of the
oscillations of roAp stars; that may have important consequences for
the study, based on seismic data, of atmospheres of roAp stars.

\subsection{Pulsation in hot subdwarf and white dwarf stars}

Pulsating hot subdwarf stars include the V1093\,Her variables,
subdwarf B (sdB) stars exhibiting gravity mode-pulsations, V361\,Hya
variables, sdB stars exhibiting pressure-mode pulsations, DW\,Lyn
variables, hybrids showing both V1093\,Her and V361\,Hya type
variability, and a unique subdwarf O pulsator.  Pulsating white dwarf
stars include the GW\,Vir variables, the hottest white or pre-white
dwarfs.  Among cooler stars they also include the helium-rich
V777\,Her (DBV) stars, the carbon-rich hot DQV pulsators and the
classical hydrogen-rich ZZ\,Ceti (DAV) stars.

Among hot subdwarfs the first {\em Kepler} survey found only one
V361\,Hya variable in the field, but several V1093\,Her variables were
identified ({\O}stensen et al.~2010). Those discoveries spawned an
industry of more detailed analyses and follow-up surveys (e.g. Reed et
al.~2010, Pablo et al.~2011, and references therein).

Detailed asteroseismic modelling for a number of sdB stars has mostly
been based on the more mature CoRoT data, although {\em Kepler} data
are also now having an impact (e.g. Charpinet et al.~2011, and
references therein). {\em Kepler} has had less influence in the
pulsating white~dwarf arena; however, representatives of the ZZ\,Cet
and V777\,Her classes in the {\em Kepler} field (Hermes et al.~2011,
{\O}stensen et al.\ 2011) have now been found.

Larger-scale surveys continue to find more variables in {\it all} of
the classes described above. The recent literature contains too many
examples of surveys to be listed individually; it is the sum of
contributions, rather than any individual publication, which makes
this a progressive field of research. The possibility that any of
these discoveries might represent something really new is investigated
in follow-up observations of individual stars. Two pulsating sdB
stars have attracted particular attention: CS\,1246 is a radial
pulsator in a close binary; fortnightly variations in the ephemeris
are caused by light-travel time delays as the star orbits a
0.12\,M$_{\odot}/\sin i$ companion (Barlow et al.~2011). Long-period
variability has been confirmed in the sdB star LS\,IV$-14^{\circ}116$
(Green et al.~2011), an object that is chemically extremely peculiar
(Naslim et al.~2011). It has been suggested that the pulsations may be
excited by the $\epsilon$ mechanism in He-burning shells
(Miller Bertolami et al.~2011), possibly making it the first pulsator
known to be excited in this way.

The asteroseismic properties of white dwarfs of all four types (DAV,
DBV, DQV and DOV) have been explored, with the DBV prototype GD 358
coming under close scrutiny (e.g.~Montgomery et al.~2010), partly in
the wider quest to characterise convection physics in DAV and DBV
white-dwarf atmospheres using non-linearities in the light curves.
Evidence that the DOV prototype GW\,Vir rotates as a solid body
(Charpinet et al.~2009) addresses the long-standing angular-momentum
question for white dwarfs: angular momentum must be removed at an
earlier phase of evolution.

\section{Catalogues and data archives}

The flow of new variable-star discoveries makes compilation of new
Name-Lists in the GCVS system a very complicated task. Thus, the 80th
Name-List will contain more than 6000 new variables and is being
published in three parts (part 1: published, part 2: end 2011, part 3:
2012). Samus et al.\ (2009) have published a catalogue of accurate
equatorial coordinates for variable stars in globular clusters.

The Div.\ V WG for Spectroscopic Data Archives was wound up in 2010 
March. In operation since 1992, it had accomplished -- or witnessed -- 
some valuable changes in attitudes towards archiving observations of 
spectra.  Having fulfilled its prime objective in those achievements, a 
new bottleneck in the form of creating archives of {\it reduced} spectra 
was recognised, in particular for echelles. Efforts to reorient the WG 
towards ``Pipeline Reductions" did not have the right audience or the 
right platform for adequate revitalization. The mission to design a 
dependable future for astronomy's heritage of spectra on photographic 
plates has been actively taken over by the Task Force for the 
Preservation and Digitization of Photographic Plates (PDPP, Comm. 5).

\section{Projected future of the Commission and its science}

Currently, Commission 27 is dominated by people working on stellar
pulsation. Given the coming changes in the structure of the IAU, we
need to consider whether to keep the current comprehensive Commission,
albeit with a better balance between its different fields, or to
propose a subdivision following the nomenclature of variable stars.

Concerning science, we are awaiting the launch (foreseen in 2013) 
of the Gaia mission. It will provide astrometry, photometry, 
spectrophotometry and spectroscopy of $\sim 10^9$ objects with $6<V<20$ 
including variable stars, over a projected mission length of 5 years, 
with an average number of measurements of $\sim 70$ per object. 
BRITE-Constellation comprises six nanosatellites to be launched 
sequentially in 2012 - 2013, aiming at variable stars with $V<4$. It 
will be the first using at least two photometric filters.  Regrettably, 
the PLATO mission has not been selected for implementation in the near 
future.

The VISTA ESO Public Survey (Minniti et al.~2010) will provide
information on about $10^6$ variable stars in several infrared
passbands, and going several magnitudes deeper than 2MASS. The survey
totals $\sim$1929 hours of observations, spread over $\sim$5 years,
and includes $\sim$35 known globular clusters and hundreds of open
clusters towards the Galactic bulge and an adjacent part of the
disk. The (predominantly) spectroscopic SONG global telescope network
is still progressing, and aims to make precise radial-velocity
measurements with a strong focus on investigating solar-like oscillations.

\section*{Acknowledgments}
The OC thanks Elizabeth Griffin and Nikolai N. Samus for their 
contributions.

\vspace{3mm}
 
{\hfill Gerald Handler}

{\hfill {\it President of the Commission}}

\end{document}